\documentclass[a4paper,12pt]{article}
\usepackage{latexsym,epsfig,mathrsfs,subfigure,psfrag,a4wide}
\renewcommand{\theequation}{\arabic{section}.\arabic{equation}}
\newcommand{\etal}{\emph{et al.}}

\newcommand{\bea}{\begin{eqnarray}}
\newcommand{\eea}{\end{eqnarray}}
\newcommand{\be}{\begin{equation}}
\newcommand{\ee}{\end{equation}}
\newcommand{\pkt}{\; .}
\newcommand{\kma}{\; ,}
\newcommand{\eqn}[1]{(\ref{#1})}

\newcommand{\calm}{{\cal M}}

\newcommand{\bfk}{{\bf k}}
\newcommand{\bfp}{{\bf p}}
\newcommand{\bfq}{{\bf q}}

\begin{document}

\begin{titlepage}
\begin{flushright}
\texttt{DO-TH-02/17}\\
October 2002\\
revised January 2003
\end{flushright}

\vspace{20mm}
\begin{center}
{\Large \bf The $O(N)$ linear sigma model at finite
temperature beyond the Hartree approximation  }

\vspace{10mm}

{\large J\"urgen Baacke \footnote{
e-mail:~\texttt{baacke@physik.uni-dortmund.de}}} and {\large Stefan Michalski
\footnote{e-mail:~\texttt{stefan.michalski@uni-dortmund.de}} } \\
\textit{ Institut f\"ur Physik, Universit\"at Dortmund} \\
\textit{D--44221 Dortmund, Germany}
\\
\vspace{8mm}
\end{center}

\begin{abstract}
\noindent
We  study  the $O(N)$ linear sigma model with spontaneous
symmetry breaking, using a Hartree-like ansatz
with a classical field and variational masses. We go beyond the Hartree 
approximation by including the two-loop contribution, the sunset diagram,
using the 2PPI expansion.
We have computed numerically the effective potential at finite temperature.
We find a phase transition of second order, while it is first order
in the one-loop Hartree approximation.
We also discuss some implications of the fact that in this order,
the decay of the sigma into two pions affects the thermal diagrams.
\end{abstract}

\end{titlepage}

\section{Introduction}
The $O(N)$ linear sigma model has a long-standing history, in particular
as a basic model for a quantum field theory with spontaneous symmetry breaking
\cite{Kirzhnits:1974as,Coleman:1974jh,Dolan:1974qd,Bardeen:1983st}.
Early investigations beyond the classical level have been based on
including one-loop quantum and thermal corrections.
These studies have been centered around the discussion of the
one-loop effective potential $V_\mathrm{eff}(\phi)$ where $\phi$ is the
mean value of the quantum field $\Phi$, in a sense being
defined more precisely by the effective action formalism,
summing up one-particle irreducible (1PI) graphs.
A next class of approximations include bubble resummations, as
motivated by the large-$N$ limit. In the model with spontaneous
symmetry breaking one finds a second order phase transition
such that the symmetry is restored at high temperature.

Another approximation, going somewhat beyond \textnormal{the leading order
of the large-$N$ expansion}, is the Hartree approximation;
\textnormal{it includes only local one-loop corrections to the effective mass
and thereby takes into account \emph{some}, but not all, next-to-leading
order corrections in $1/N$. }
The Hartree approximation of the $O(N)$ linear sigma model has been 
studied at finite temperature by various authors
\cite{Amelino-Camelia:1993nc,Amelino-Camelia:1997dd,Petropoulos:1998gt,
Nemoto:1999qf,Lenaghan:2000si,Patkos:2002xb,Verschelde:2000ta}; 
the model with spontaneous symmetry
breaking is found to have a phase transition of first order towards
the symmetric phase at high temperature.
 In contrast to the large-$N$ case the mass of
the pion quantum fluctuations does not vanish in the broken phase.
This has been discussed  as a ``violation of the Nambu-Goldstone theorem''
\textnormal{(see e.g. Ref.~\cite{Amelino-Camelia:1997dd} and references therein)};
the presently accepted point of view 
\textnormal{\cite{Nemoto:1999qf,Verschelde:2000ta,vanHees:2001ik}}
is that the ``sigma and pion masses''
in the Hartree scheme are just  variational parameters, and not
the real pion and sigma masses,
which are to be computed from the effective potential at its minimum.

If one wants to go beyond the large-$N$ and Hartree approximations
there is a variety of choices. \textnormal{The systematic expansions are 
based on the resummation scheme by Cornwall, Jackiw and Tomboulis (CJT) 
\cite{Cornwall:1974vz}, the 2PI scheme. Within this scheme one may
select certain groups of graphs in order to obtain systematic
expansions in $1/N$ or in the number of loops (order in $\hbar$).
Beyond the leading order, these extensions require technically quite 
involved analytical and
numerical calculations \cite{vanHees:2001ik}.} In general one has 
 to solve Schwinger-Dyson equations for the Green functions which 
in the present case would even form a coupled system.
\textnormal{Little is known about the merits of the next-to-leading order
extensions as such calculations at finite temperature in
$3+1$ dimensions are not yet available.}

A technically less demanding approach is the 2PPI resummation introduced
by Verschelde \cite{Verschelde:2PPI,Verschelde:2000dz}. 
Here, instead of treating the Green functions as variational parameters 
one just introduces variational masses, like in the Hartree approximation. 
This implies that the resummation
is only over local insertions, the 2-particle ``point reducible''
graphs, i.e., graphs that fall apart if one cuts two lines meeting at the same
point (the 2PPR point). 
\textnormal{This approach is based on a variational principle for expectation
values of local composite operators, i.e., all of the system's
equations of motion can be derived from a single functional.} 
The 2PPI effective action is identical to that in the  Hartree approximation
if only one-loop 2PPI graphs are included; this has been studied in
Ref. \cite{Verschelde:2PPI, Verschelde:2000ta}. 
For the complete two-loop approximation
one must also include the sunset diagram.
For the case $N=1$ Smet \etal\
\cite{Smet:2001un} have evaluated the effective potential; they found that
instead of a first order phase transition one obtains a second order one.
Here we extend this investigation to the case of general $N$.

The $O(N)$ linear sigma model with spontaneous symmetry breaking
 has been studied in nonequilibrium quantum
field theory as well, mostly in the large-$N$ limit
\textnormal{and with different initial conditions for the mean field
$\phi=\langle\Phi\rangle$ and for the density matrix of the fluctuations}
\cite{Cooper:1997ii,Boyanovsky:1999yp,
Destri:2000hd,Borsanyi:2000pm,Borsanyi:2002tm,
Felder:2000hj,Felder:2001kt,Garcia-Bellido:2002aj,Baacke:2000fw}.
\textnormal{There} the mean field $\phi=\langle\Phi\rangle$ becomes
time dependent. As far as symmetry  restoration is concerned striking
similarities with finite temperature quantum field theory are
observed \textnormal{\cite{Cooper:1997ii,Boyanovsky:1999yp,Baacke:2000fw}:}
  If the system is supplied with a high initial energy density
it displays symmetry restoration at late times
in the sense that the mean field settles at $\phi=0$ or oscillates
around this value, while at lower energy densities the system ends up in a
broken symmetry phase where the time average of $\phi(t)$ remains different
from zero, and where the pion mass, the time-dependent mass of the
quantum fluctuations, goes to zero. This \textnormal{phase structure}
persists if one uses the
Hartree instead of the large-$N$ approximation \cite{Baacke:2001zt};
\textnormal{in this case, as in thermal equilibrium, the effective mass of
the pion fluctuations remains finite even at low energy densities.}

However, in the large-$N$ or Hartree approximations the system does
not approach thermal equilibrium. This problem has been addressed
in a general way  in Ref. \cite{Ramsey:1997qc,Calzetta:2002ub}.
Numerous authors
\cite{Berges:2000ur,Berges:2001fi,Mihaila:2000ib,Mihaila:2001sr,
Blagoev:2001ze,Aarts:2002dj,Cooper:2002ze,Cooper:2002qd} have
tried recently to find useful approximations beyond the
leading orders. Up to now numerical simulations
are mostly  limited to $1+1$ dimensional models.
Most of the new approximations show large deviations from
the large-$N$ approximation, and they indicate thermalization.
The proper case of an $O(N)$ model with spontaneous symmetry breaking
has not been investigated up to now. Indeed  higher corrections
have not even been included in equilibrium calculations for such models.
If one tries to appreciate the quality of various approximations
such equilibrium computations should be able to yield useful
additional insights. It is one of the purposes of this work to
initiate such investigations.

The plan of the paper is as follows:
 in section~\ref{sec:basic eq} we present the general formulation 
of the model and
of the 2PPI formalism.
In section~\ref{sec:computation} we  explicitly formulate a potential
$U(m_\sigma^2,m_\pi^ 2,\phi)$ that by variation of $m_\sigma^2$ and
$m_\pi^2$ leads to the gap equations. The technical details of the
relevant Feynman graphs and a comparison of the 2PPI expansion to CJT's
2PI approach are presented in the appendices.
In section~\ref{sec:discussion} we discuss our numerical results, we end with
a summary and an outlook in section~\ref{sec:conclusions}.

\section{Basic equations}
\label{sec:basic eq}
\setcounter{equation}{0}
The Lagrange density of the $O(N)$ linear sigma model is given by
\be
\label{eq:Lagrange}
 \mathscr{L} =\frac{1}{2}\partial_\mu\Phi_i\,\partial^\mu\Phi_i
-\frac{\lambda}{4}\left(\Phi_i\Phi_i-v^2\right)^2
\kma
\ee
where $\Phi_i$ is a vector with $N$ components.
We intend to compute the effective potential of this model at finite
temperature. This model has been studied at large $N$ and in
the Hartree approximation, which both represent bubble resummations.
One of the possibilities
to go beyond these approximations, and in particular to include higher loop
corrections is the use of the 2PI or CJT formalism; this is technically
involved, even in equilibrium, as one has to solve Schwinger-Dyson
equations for the Green functions, in the present case indeed a coupled system
of integral equations.

Another possibility of going beyond the leading order approximations
has been proposed by Verschelde \cite{Verschelde:2PPI,Verschelde:2000dz}, 
the so-called 2PPI formalism.  
\textnormal{This is a variant of the 2PI (CJT) formalism by Cornwall, Jackiw
and Tomboulis~\cite{Cornwall:1974vz}. In the 2PPI approach the
composite operator $\Phi_i\Phi_j$ is \emph{local} while in 2PI
it is \emph{bilocal}. }
Here the resummation encompasses all 2-particle
{\em point reducible} graphs,
graphs that fall apart if two lines meeting at one point (vertex), 
the 2PPR point, are cut.
These graphs are deleted in the 1PI effective action, which thereby
is replaced by the 2PPI effective action. They are taken into account
by a mass insertion like in the Hartree approximation --- to which the 2PPI
expansion reduces in the one-loop approximation.
\textnormal{ We compare this approach to the well-known 2PI CJT formalism in 
Appendix~\ref{sec:2PI_2PPI}.}

The problem occurring  in the Hartree approximation, namely
the lack of a consistent renormalization, has been solved in a systematic way.
The inconsistencies are avoided by recognizing that in the resummation
the counterterms have to be divided into 2PPI and 2PPR parts.
The 2 particle point reducible parts renormalize the gap equation, the
2PPI parts renormalize the 2PPI effective action.
This procedure has been discussed in technical detail
in \cite{Verschelde:2000dz,Verschelde:2000ta} and
been applied to a first two-loop
calculation for the $N=1$ model \cite{Smet:2001un}, including the
sunset diagram as the only 2-loop 2PPI term, the only other new
terms being one-loop
graphs computed with the one-loop counterterm Lagrangian.

We will not go into details here. For the $O(N)$ case we use the
explicit formulae of \cite{Verschelde:2000dz}.
The classical field is denoted by $\phi_i=\langle\Phi_i\rangle$, the
 bubble resummation is defined by introducing
local insertions $\Delta_{ij}=\langle\Phi_i\Phi_j\rangle-\phi_i\phi_j=
\langle\Phi_i\Phi_j\rangle_{conn.}$ which 
collect all 2PPR graphs.
The resummation is defined by including these insertions as well
as the seagull insertions, it is obtained by introducing into the
propagators the effective mass
\footnote{Our convention for the coupling $\lambda$ differs
from the one in Ref.\cite{Verschelde:2000dz} by a factor of 2.}
\be
\overline m_{ij}^2=-\lambda v^2 \delta_{ij}+
2\lambda\left[(\phi_i\phi_j+\Delta_{ij}\right)+
\lambda\left(\phi_k\phi_k + \Delta_{kk})\right]
\pkt\ee

\textnormal{The motivation and formal derivation of the 2PPI effective action
for the case $N=1$ is presented in Appendix~\ref{sec:2PI_2PPI}.
The generalization to arbitrary $N$ is straightforward 
\cite{Verschelde:2000ta}.
The 2PPI effective action can be written as}
\be
\Gamma=S_\mathrm{class}+\Gamma^\mathrm{2PPI}_q[\phi_i,\overline{m}^2_{ij}]
-\frac{\lambda}{4}\left(\Delta_{ii}\Delta_{jj}+2\Delta_{ij}\Delta_{ij}\right)
\pkt\ee
It includes all 2PPI graphs as defined above,
with the mass terms replaced by the variational masses $\overline m_{ij}$,
and it is computed using the 2PPI parts of the counterterms.
The last term is introduced in order to avoid double counting.
\textnormal{The local self-energies $\Delta_{ij}$}
 can be shown to be related to 
\textnormal{the ``quantum'' part of the 2PPI action via}
\be
\frac{1}{2}\Delta_{ij}=\frac{\partial \Gamma_q^\mathrm{2PPI}(\overline m_{ij}^2)}
{\partial \overline m_{ij}^2}
\ee
which defines a self-consistency condition or gap equation.

For $\Delta_{ij}$ and $\overline m_{ij}$ one uses the $O(N)$ invariant
\emph{ans\"atze}
\bea
\overline m_{ij}^2&=&\frac{\phi_i\phi_j}{\phi^2}m_\sigma^2+
\left(\delta_{ij}-\frac{\phi_i\phi_j}{\phi^2}\right)m_\pi^2
\kma \\
\Delta_{ij}&=&\frac{\phi_i\phi_j}{\phi^2}\Delta_\sigma+
\left(\delta_{ij}-\frac{\phi_i\phi_j}{\phi^2}\right)\Delta_\pi
\kma \eea
so that the equations for the effective masses
separate as
\bea \nonumber
m_\sigma^2&=&\lambda\left[3\phi^2-v^2+3\Delta_\sigma
+(N-1) \Delta_\pi\right] \kma
\\ \label{gaps1}
m_\pi^2&=&\lambda\left[\phi^2-v^2+\Delta_\sigma
+(N+1) \Delta_\pi\right]
\pkt \eea
The gap equations become
\bea\nonumber
\frac{\delta \Gamma_q^\mathrm{2PPI}}{\delta m_\sigma^2}&=&\frac{1}{2}\Delta_\sigma
\kma
\\ \label{gaps2}
\frac{\delta \Gamma_q^\mathrm{2PPI}}{\delta m_\pi^2}&=&\frac{1}{2}(N-1)\Delta_\pi
\kma\eea
and the effective potential takes the form
\bea \nonumber
V_\mathrm{eff}(m_\sigma^2,m_\pi^2,\phi)&=&
\frac{\lambda}{4}(\phi^2-v^2)^2+V_q^\mathrm{2PPI}(m_\sigma^2,m_\pi^2,\phi)
\\ \label{V1PI}
&&-\frac{\lambda}{4}\left(3\Delta_\sigma^2+(N^2-1)\Delta_\pi^2+
2(N-1)\Delta_\sigma\Delta_\pi\right)\ ,
\eea
\textnormal{where $V_q^\mathrm{2PPI}$ is the ``quantum'' part of the 2PPI effective
potential.}
As has been shown in Refs. \cite{Verschelde:2000ta,Verschelde:2000dz} these
equations can be properly renormalized and the renormalized equations
have the same form.  We do not discuss this here.
For the numerical calculation we have used the renormalized versions of these
equations; we have not put renormalization conditions but used an
$\overline{\rm MS}$ prescription. The renormalization scale $\overline \mu$
refers to this prescription.

\section{Computation of the effective potential}
\label{sec:computation}
\setcounter{equation}{0}
The basic relations given in the previous section can be used to compute
the effective potential. We would have to solve the coupled
system of gap equations and to insert the result into the
1PI effective action. This would imply that we would not only have
to evaluate the sunset graphs, but also their derivatives with respect
to $m_\sigma^2$ and $m_\pi^2$.
Here we prefer to work with an effective potential that leads
to the gap equations by finding the extremum (maximum) with respect
to variations of $m_\sigma^2$ and $m_\pi^2$. Instead of
solving the gap equations whose algebraic and analytic form is
already quite involved, we then can simply use numerical algorithms
for extremizing a function of two variables \cite{NumRec}.
\textnormal{To this end we solve  Eqs.~\eqn{gaps1} with respect to 
$\Delta_\sigma$ and
  $\Delta_\pi$ and insert these expressions into Eq.~\eqn{V1PI}. 
We denote this new potential by $U(m_\sigma^2,m_\pi^2,\phi)$.
It can easily be verified, that Eqs.~\eqn{gaps1} again follow
by extremizing this potential with respect to $m_\sigma^2$ and $m_\pi^2$.
The 1PI effective potential as a function of $\phi$ alone is obtained as
\be
V^{1PI}_\mathrm{eff}(\phi)=U(\overline m_\sigma^2,\overline m_\pi^2,\phi)
\ee
where $\overline m_ \sigma$ and $\overline m_\pi$ are the values which
extremize (maximize) $U$ for a given $\phi$.} 
This procedure, as introduced by Nemoto
\etal\  \cite{Nemoto:1999qf} in the Hartree approximation,  generalizes
to the case where higher order contributions are included into
$\Gamma^\mathrm{2PPI}_q$ (see also Appendix~\ref{sec:2PI_2PPI}.2). 
Here we include the two-loop contribution,
the sunset diagram, as has been done previously for the
$N=1$ model by Smet \etal\ \cite{Smet:2001un}.

With these preliminaries we can now give our explicit equations:
We decompose the potential $U(m_\sigma^2,m_\pi^2,\phi)$ into three parts:
\be
U=U_\mathrm{class} + U_\mathrm{1-loop} + U_\mathrm{sunset}
\pkt
\ee
The classical potential has the form (see \cite{Nemoto:1999qf})
\bea
U_\mathrm{class}&=&\frac{1}{2} m_\sigma^2\phi^2-\frac{\lambda}{2} \phi^4 -
\frac{1}{2\lambda (N+2)}v^2\left\{m_\sigma^2 +(N-1)m_\pi^2\right\}
\\ \nonumber
&&-\frac{1}{8\lambda(N+2)}\left[
(N+1)m_\sigma^4+3(N-1)m_\pi^4-2(N-1)m_\sigma^2m_\pi^2+2 N \lambda^2 v^4\right]
\kma\eea
one  easily checks that it takes its {\em maximum}
if $m_\sigma^2=\lambda(3\phi^2-v^2)$ and
$m_\pi^2=\lambda(\phi^2-v^2)$.
The 1-loop part is given by the ``ln det'' contributions. 
\textnormal{At finite temperature these include the free energies, so the
one-loop part of the effective action reads}
\bea \nonumber
U_\mathrm{1-loop}&=&
\frac{1}{2}\int\frac{d^4 k}{(2\pi)^4}\ln(k^2+m_\sigma^2)
+\frac{N-1}{2}\int\frac{d^4 k}{(2\pi)^4}\ln(k^2+m_\pi^2)
\\
&&+T\int\frac{d^3 k}{(2\pi)^3}\ln\left[1-\exp(-E_\sigma(\bfk)/T)\right]
\\\nonumber&&
+ (N-1)T\int\frac{d^3 k}{(2\pi)^3}\ln\left[1-\exp(-E_\pi(\bfk)/T)\right]
\pkt\eea
\textnormal{In computing the sunset diagram it is convenient to decompose
the finite temperature propagators into a zero temperature and
a finite temperature (thermal) part
proportional to $\delta(k^2-m_j^2)/\exp(-E_j/k_BT)$.}
The contribution of the sunset diagrams then consists of three parts
(see e.g. Ref. \cite{Parwani:1991gq}) 
\be
U_\mathrm{sunset}=U^{(0)}_\mathrm{sunset}+ U^{(1)}_\mathrm{sunset}+U^{(2)}_\mathrm{sunset}\ ,
\ee
with the $T=0$ contribution
$U_\mathrm{sunset}^{(0)}$, the diagrams with one thermal line $U_\mathrm{sunset}^{(1)}$ and
the diagrams with two thermal lines $U^{(2)}_\mathrm{sunset}$, 
\textnormal{see Fig.~1.}
\begin{figure}[htbp]
  \centering
  \psfragscanon
    \psfrag{s}{\scriptsize $\sigma$}
    \psfrag{p}{\scriptsize $\pi$}
    \psfrag{9}{\scriptsize $9$}
    \psfrag{3}{\scriptsize $3$}
    \psfrag{\(N-1\)}{\scriptsize $(N-1)$}
    \psfrag{2\(N-1\)}{\scriptsize $2(N-1)$}
    \psfrag{+}{\scriptsize $+$}
    \psfrag{9}{\scriptsize $9$}
  \subfigure[Zero temperature part.]
  {\includegraphics[scale=0.45]{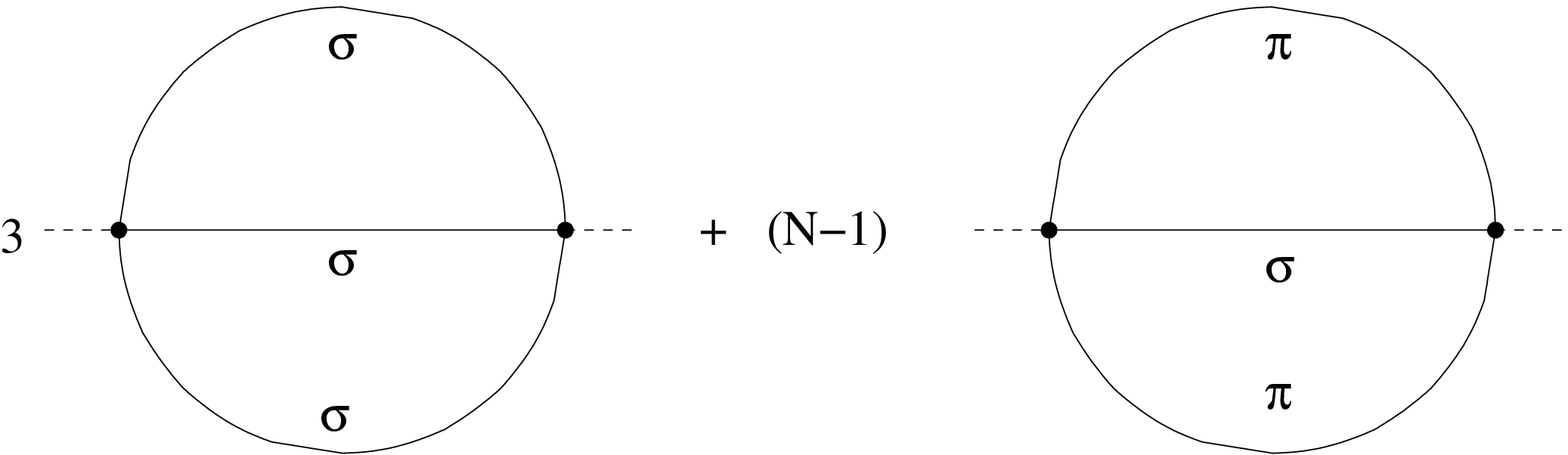}}
  \subfigure[Graphs with one thermal line.]
  {\includegraphics[scale=0.45]{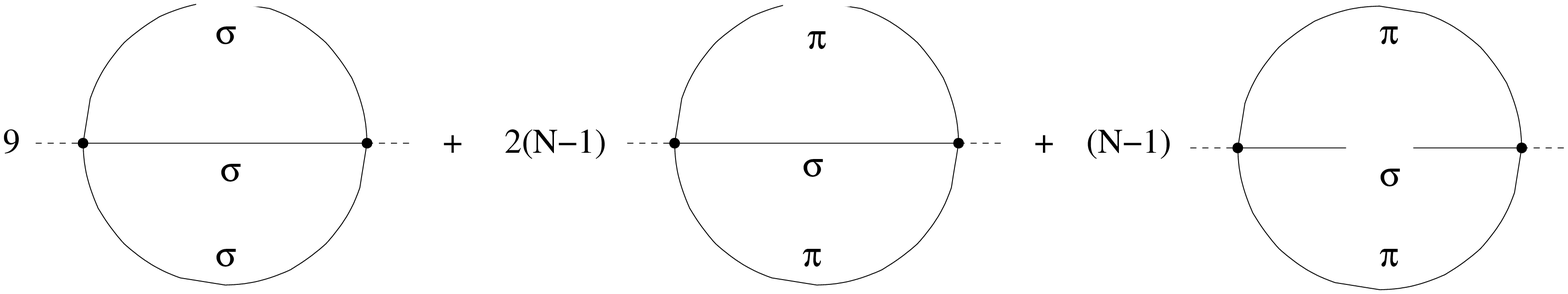}}
  \subfigure[Graphs with two thermal lines.]
  {\includegraphics[scale=0.45]{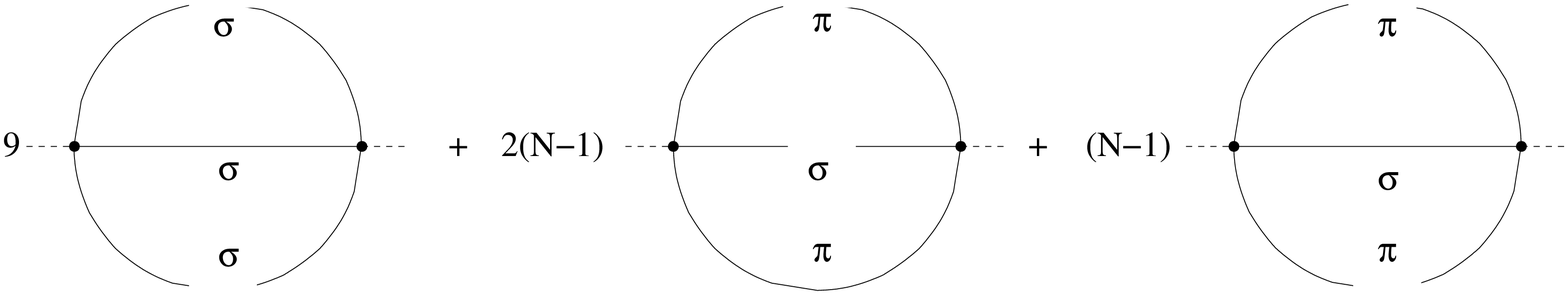}}
  \caption{Contributions to the sunset diagram at finite temperature. 
    Solid lines represent the zero temperature parts of the propagators,
    finite temperature parts are denoted by interrupted lines.}
  \label{fig:1}
\end{figure}
The $T=0$ part is given by
\be
U^{(0)}_\mathrm{sunset}=-\lambda^2\phi^2
\left[3 I_{\sigma\sigma\sigma} + (N-1) I_{\sigma\pi\pi}\right]
\kma\ee
it is represented graphically in Fig.~\ref{fig:1}a.
The diagrams with one thermal line, see Fig.~\ref{fig:1}b, contribute
\be
U^{(1)}_\mathrm{sunset}=-\lambda^2\phi^2
\left[9 I^\beta_{\sigma\sigma|\underline{\sigma}} +
(N-1) \left(2I^\beta_{\sigma\pi|\underline{\pi}}
+I^\beta_{\pi\pi|\underline{\sigma}}\right)\right]
\pkt\ee
The symbols for the thermal lines are underlined.
Similarly the diagrams with two thermal lines, see Fig.~\ref{fig:1}c,
contribute
\be
U^{(2)}_\mathrm{sunset}=-\lambda^2\phi^2
\left[9 I^\beta_{\sigma|\underline{\sigma}\underline{\sigma}} +
(N-1) \left(2I^\beta_{\pi|\underline{\sigma}\underline{\pi}}
+I^\beta_{\sigma|\underline{\pi}\underline{\pi}}\right)\right]
\pkt\ee
The precise definition of the Feynman integrals
$I_{ijk}$~,~$I^\beta_{ij|\underline k} $ and
$I^\beta_{i|\underline{jk}}$ with zero, one and two
thermal lines, respectively, as well as their analytic form
are presented in the Appendices. It is understood that their divergent parts
are removed.

\section{Discussion of the numerical results}
\label{sec:discussion}
\begin{figure}[htbp]
   \centering
   \psfrag{T}{$T$}
   \psfrag{v(T)}{$v(T)$}
   \includegraphics[scale=0.4]{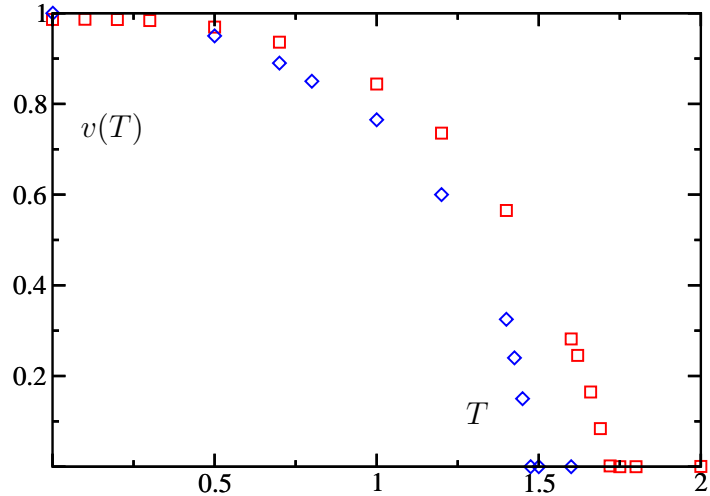}
   \psfragscanoff
   \caption{The expectation value $v(T)$ of $\phi$ for $N=4$, 
     $\bar{\mu}=v=1$, $\lambda=1$ (squares) and $\lambda=0.1$ (diamonds).}
   \label{fig:2}
 \end{figure}

\setcounter{equation}{0}
As we have stated previously we do not solve the two coupled gap equations
but instead we maximize the potential $U(m_\sigma^2,m_\pi^2,\phi)$.
We present our numerical  results for the case $N=4$ with
$\lambda=1$ and $\lambda=0.1$ 
\footnote {The choice $\lambda=1$ corresponds to $\lambda=2$ in the
normalization of Ref.~\cite{Verschelde:2000ta} and to $\lambda=6$
when compared to the $N=1$ study of Ref. \cite{Smet:2001un}}.
The mass scale is fixed by taking
$v=1$, and we choose the $\overline{\rm MS}$ renormalization
scale $\overline \mu=1$. In Fig.~\ref{fig:2} we display the value
of $\phi$ at the minimum
of the effective potential, the thermal expectation value
 which we denote by $v(T)$.
If we choose $\lambda=1$ we see a phase transition 
towards the symmetric phase  $v(T)=0$ for $T>T_0$ with $T_0\simeq 1.7$.
For $\lambda=0.1$ the critical temperature is about $T_0\simeq1.475$.

\begin{figure}[htbp]
   \centering
   \psfragscanon
   \psfrag{V}{$V_\mathrm{eff}(\phi)$}
   \psfrag{eff }{}
   \psfrag{\( }{}
   \psfrag{f \)}{}
   \psfrag{f}{$\phi$}
   \subfigure[ $\lambda=1$ for temperatures
     $T=1.62,1.66,1.69,1.70$ and $1.72$]
     {\includegraphics[scale=0.4]{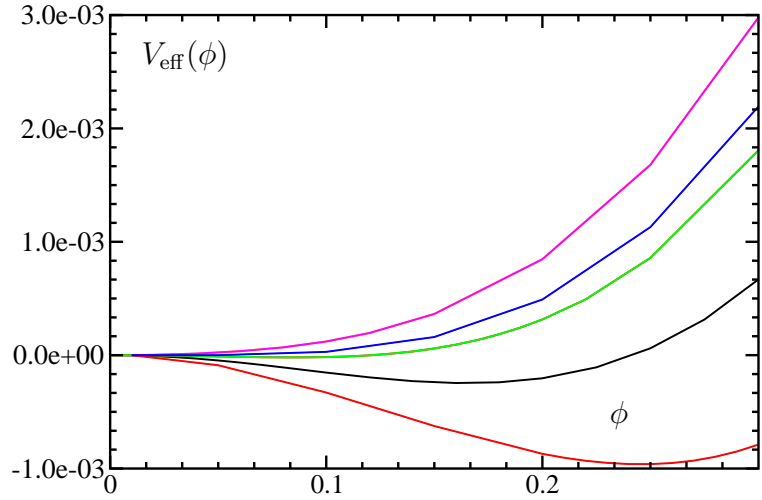}}
   \subfigure[$\lambda=0.1$ for temperatures
     $T=1.2,1.4,1.425,1.45,1.475,1.5$ and $1.6$.]
   {\includegraphics[scale=0.4]{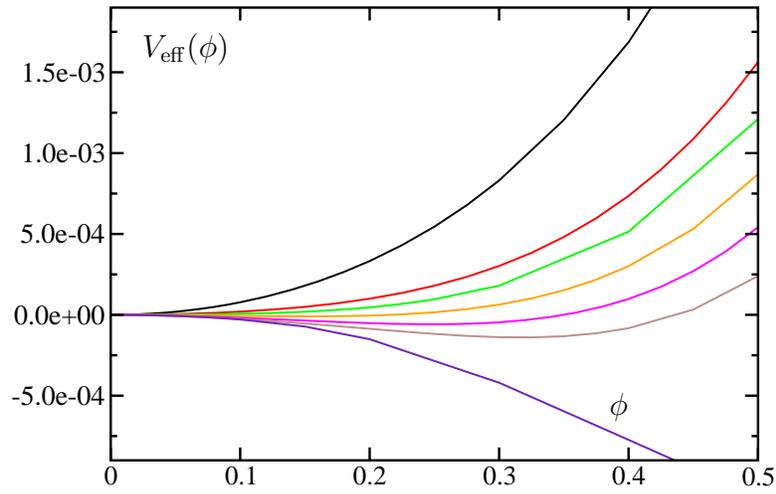}}
   \psfragscanoff
   \caption{The behavior of $V_\mathrm{eff}(\phi)$ near the critical 
     temperature, parameter set as in Fig.~\ref{fig:2}.}
   \label{fig:3}
 \end{figure}

\begin{figure}[htbp]
  \centering
  \psfragscanon
  \psfrag{V}{$V_\mathrm{eff}(\phi)$}
  \psfrag{eff}{}
  \psfrag{\(}{}
  \psfrag{f \)}{}
  \psfrag{f}{$\phi$}
  \includegraphics[scale=0.4]{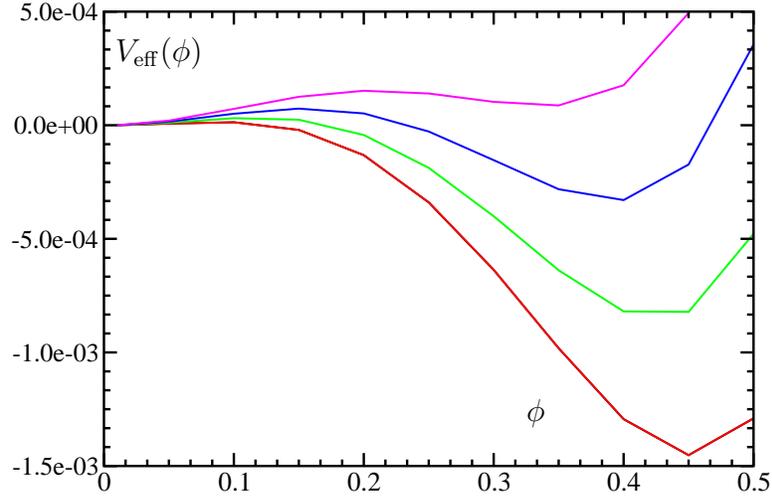}
  \psfragscanoff
  \caption{The behavior of $V_\mathrm{eff}(\phi)$ near the critical temperature
in the Hartree approximation; the curves are for $T=1.46,1.47,1.48$ and $1.49$;
parameter set as in Fig.~\ref{fig:2} with $\lambda=1$.}
  \label{fig:4}
\end{figure}

The behavior of the effective potential as a function of $\phi$
near the phase transition is displayed in Fig.~\ref{fig:3},
which clearly indicates that the phase transition is of second order.
The behavior, for the same parameters ($\lambda=1$) but without the 
sunset diagram (i.e. Hartree approximation, see Ref.~\cite{Smet:2001un}),
is displayed in Fig.~\ref{fig:4}. The two minima characteristic of a
first order phase transition are well visible.
It is well known that a phase transition of first
order is found in the Hartree approximation. As apparent from the
scale on the $y$ axes and from the tiny temperature
range, Figs. 3 and 4 represent  ``microscopic'' pictures of the
two phase transitions.

\begin{figure}[htbp]
  \centering
  \psfragscanon
  \psfrag{M}{$M_\sigma$}
  \psfrag{T}{$T$}
  \psfrag{s}{}
  \includegraphics[scale=0.4]{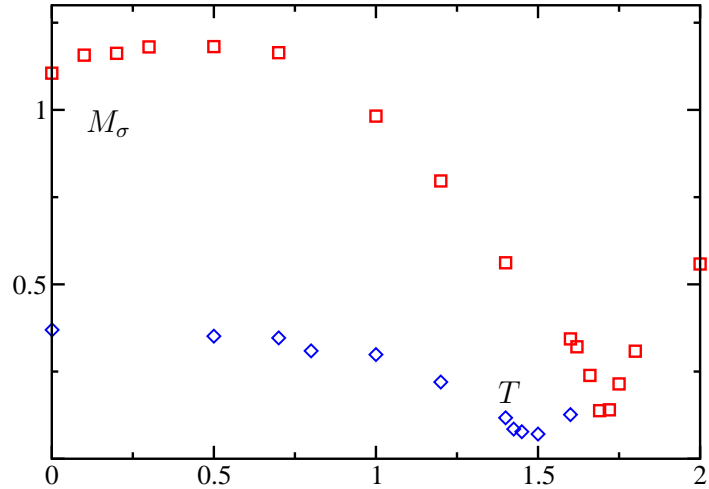}
  \caption{The sigma mass $M_\sigma$ obtained from the effective potential
 as a function of temperature; parameter sets as for Fig.~\ref{fig:2}}
  \label{fig:5}
\end{figure}

The temperature dependence of the sigma mass $M_\sigma$ as defined by the
curvature of the effective potential at its minimum is shown
in Fig.~\ref{fig:5}. As to be expected it goes to zero at the phase transition
temperature, the zero is approached linearly if one plots $M_\sigma^2$.

\begin{figure}[htbp]
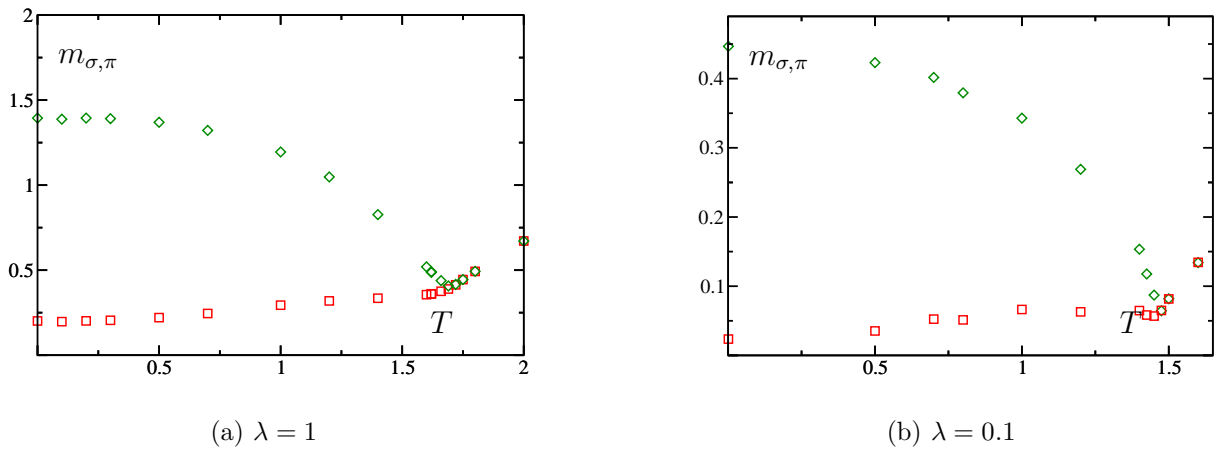

  \centering
  \psfrag{m}{$m_{\sigma,\pi}$}
  \psfrag{s,p}{}
  \psfrag{T}{$T$}
  \psfrag{s}{}
  \subfigure[$\lambda=1$]{
  \includegraphics[scale=0.3]{figure6a.eps}}
  \hfill
  \subfigure[$\lambda=0.1$]{
  \includegraphics[scale=0.3]{figure6b.eps}}
  
  \caption{The variational masses $m_\sigma$ (diamonds) and $m_\pi$
    (squares) as functions of temperature; parameter sets as for 
    Fig.~\ref{fig:2}.}
  \label{fig:6}
\end{figure}
In Fig.~\ref{fig:6} we display the temperature dependence of the variational
masses $m_\sigma$ and $m_\pi$ at the minimum of the effective potential,
$\phi=v(T)$. The variational sigma mass $m_\sigma$ behaves similarly as
the sigma mass $M_\sigma$ obtained from the effective potential.
The mass $m_\pi$ becomes identical to the mass $m_\sigma$ above the
phase transition, but does not vanish below the phase transition.
It was found already in the one-loop analysis that, as a ``violation
of the Nambu-Goldstone theorem'', the self-consistent pion masses
do not vanish when the symmetry is broken. It has been argued
that these self-consistent masses are not the physical pion masses;
indeed they are not: they are variational parameters and
the effective potential
$V_\mathrm{eff}(\phi)$ has of course $N-1$ flat directions at its minimum.
But the hope -- or expectation -- that the discrepancy between the
pion mass as computed from the effective potential, and the
pion mass as a variational parameter would disappear, turns out to
be fallacious. Indeed there is a simple physical reason for this
result, as will be discussed below the next paragraph.

As the thermal integrals require both masses to be real we have
looked for the extremum with respect to $m_\sigma$ and $m_\pi$.
So in our numerical approach neither $m_\sigma^2$
nor $m_\pi^2$ can get negative and
the well-known instability which occurs for $N=1$ in the
region where the potential has a negative curvature is
avoided {\em by fiat}, the maximum simply occurs at the
boundary of the ``physical region'' of real pion and sigma masses.
It has to be said, though, that in this case we do not solve the gap
equation which in fact becomes meaningless. In the large-$N$ limit
such a construction leads to an effective potential that
is flat in the region $\phi < v$, and therefore convex.
We do not want to enter into this discussion here, we simply
state that these regions require another approach and that we
have to discard them. Such regions occur at low temperatures only, and
of course they do not include the region around the minimum of the
effective potential. At higher temperatures, but well
below the phase transition
the effective potential has regions of negative curvature, but both
$m_\sigma$ and $m_\pi$, and therefore the effective potential are still
real, as the variational mass $m_\sigma$  is not equal to the curvature of
the potential.
The parameter $m_\pi$, which is imaginary below the minimum in the large-$N$
approximation, also becomes real  for all values of $\phi$ at temperatures
well below the phase transition.

However, we are faced with an even more important new feature:
the fact that the sigma can decay into two pions if $m_\sigma > 2m_\pi$.
The sunset diagram with one thermal sigma line and two pion lines
 acquires an imaginary part
in this case. In our computations we have simply omitted this imaginary part,
but obviously
we would  not be able to solve the gap equation in regions
where such a decay is possible. While we did not exclude
these regions we have to consider the {\em real part} of
the effective potential in these regions with suspicion. In contrast
to the problem of imaginary masses and the associated instability
{\em these regions do include the minimum of the effective potential
for temperatures up to almost the phase transition temperature}.
We display the (trial) masses $m_\sigma$ and $m_\pi$ at the minimum
of the effective potential in Fig.~\ref{fig:6}. Around the phase transition
itself we find  $m_\sigma < 2m_\pi$, so that the behavior
of the effective potential in the critical region, as
plotted in Fig.~\ref{fig:3}, is not affected,
but the results below $T\simeq 1.5$ for $\lambda=1$ and below $T\simeq 1.4$
for $\lambda=0.1$, respectively, 
have to be taken with some
caveat.

This finding has important consequences:
Of course an unstable
particle can coexist with its decay products at finite temperature, but
this situation requires an approach where the transitions are taken into
account; however, this is not the case in this approximation, and indeed
with the entire formalism used here.
Indeed in the regions affected by this instability our
approximation becomes inconsistent, and this should be so
\emph{a fortiori} if one considers the massless {\em physical} pions.
As the masslessness of the Goldstone particles is an important
aspect of spontaneously broken symmetry,  this
problem should be studied in detail. Of course in the applications
to real pions in the linear sigma model the pions receive
a finite mass due to explicit symmetry breaking, and the sigma particle
is considered usually as being of a problematic status anyway, hinting at
the limitations of the model as an effective theory of strong interactions.

\section{Summary and Outlook}
\label{sec:conclusions}
We have analyzed here the $O(N)$ linear sigma model in the 2PPI
formalism beyond the leading order, in which it coincides with the
Hartree approximation. As in the Hartree approximation  and in
the $N=1$ version we find that the effective mass of the pion
quantum fluctuations is different from zero in the
broken symmetry phase, so that a naive particle interpretation,
suggested by the large-$N$ analysis, becomes problematic.
As in the $N=1$ version of the model \cite{Smet:2001un}
the phase transition, which is first order in the Hartree approximation,
becomes second order. In addition to the $N=1$ case there is a new instability
associated with the possibility of the decay $\sigma \to 2 \pi$.
This will not be problematic at low temperatures and for small
couplings, but whenever the sunset diagrams become important it requires
reconsidering the entire framework. We find that near the phase transition
the sigma fluctuations become stable, as they are trivially
in the symmetric phase.

Our analysis should have some bearing on nonequilibrium simulations as well.
The non-vanishing effective mass of the ``Goldstone'' quantum
fluctuations makes it hard to maintain a naive particle interpretation;
but this is the case \emph{a fortiori} for any nonequilibrium simulations
that include higher order diagrams, for approximations in which
the propagator is not an effective free particle propagator.
In addition, however, it becomes obvious that
the additional instability that occurs only for $N >1$ will
lead to other and new aspects of such nonequilibrium simulations,
when compared to those for the large-$N$ case.
While it is certainly important to understand thermalization,
the instabilities both of the 1-loop effective potential as those
introduced by particle decay may have consequences of similar
importance, and conclusions drawn from $N=1$ simulations may
therefore miss essential aspects for models with spontaneous symmetry
breaking.

We would finally like to remark that all the complications
found in thermal equilibrium occur
in nonequilibrium studies as well, both in the preparation
of the initial state and in the analysis of the final state, as
well as in renormalization.
Therefore it is mandatory  that such equilibrium studies are being
pursued in parallel to the nonequilibrium ones.

\subsection*{Acknowledgments}
The authors take pleasure in thanking Henri Verschelde and
Andreas Heinen for useful discussions, and the Deutsche Forschungsgemeinschaft
for financial support under Ba703/6-1.  

\newpage

\begin{appendix}
\renewcommand{\theequation}{\Alph{section}.\arabic{equation}}





\section{Comparison of 2PPI and 2PI effective action}
\label{sec:2PI_2PPI}
We compare the 2PI with the 2PPI effective action using a loop expansion.
The Lagrangian of the classical action is given by Eq.~(\ref{eq:Lagrange}) 
but here we set $N=1$ for the sake of convenience. 

\subsection{2PI expansion}
\begin{figure}[htbp]
  \centering
  \psfrag{G}{$\Gamma_2^{2PI}$}
  \psfrag{2PI}{}
  \psfrag{2}{}
  \psfrag{=}{$=$}
  \psfrag{+}{$+$}
  \epsfig{file=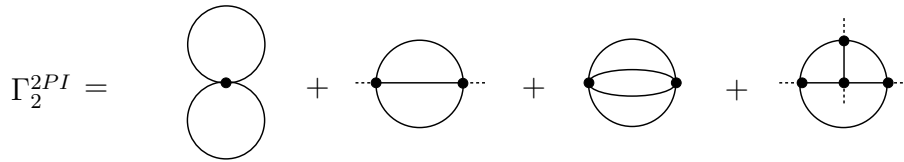,width=12cm}
  \caption{Two-loop and some (not all) three-loop contributions to the
    2PI effective action: double-bubble, sunset, basketball and another
    three-loop graph. A line represents the \emph{full} propagator
    as defined in Eq.~(\ref{eq:2PI_G}).}
  \label{fig:Gamma_2PI}
\end{figure}

The 2PI effective action reads (cf. Eq.~(2.9a) of Ref.~\cite{Cornwall:1974vz})
\begin{equation}
  \label{eq:Gamma2PI}
  \Gamma^\mathrm{2PI}[\phi,G] = S[\phi] + i\frac{\hbar}{2} \ln\det G^{-1}
  + i \frac{\hbar}{2} \mathrm{Tr}\ \mathcal{D}^{-1} G
  + \Gamma_2^{2PI}[\phi,G]\ ,
\end{equation}
where $\Gamma_2^\mathrm{2PI}$ contains all higher 2PI corrections;
the first graphs of the loop expansion are shown in Fig.~\ref{fig:Gamma_2PI}.

\begin{figure}[htbp]
  \centering
  \psfrag{+}{$+$}
  \psfrag{S}{$\Sigma(k)$}
  \psfrag{k}{}
  \psfrag{\(}{}
  \psfrag{\) =}{=}
  \epsfig{file=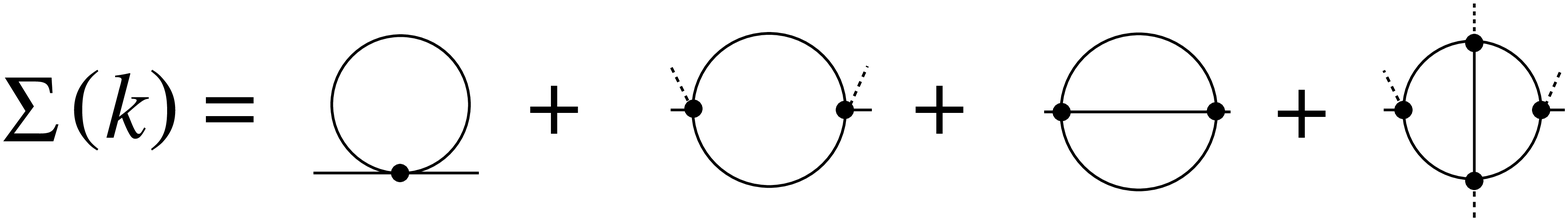,width=12cm}
  \caption{Some one- and two-loop contributions to the proper self-energy 
    of the 2PI expansion: tadpole, fish, sunset and another two-loop
    graph. A line represents a \emph{full} propagator
    as defined in Eq.~(\ref{eq:2PI_G}).}
  \label{fig:Sigma}
\end{figure}

The propagator $G$ takes the form
\begin{equation}
  \label{eq:2PI_G}
  G(k) = \frac{i}{k^2-\lambda(3\phi^2-v^2)- i\Sigma(k)}
\end{equation}
where the equation for the proper self-energy $\Sigma$ follows from the
condition $\frac{\delta\Gamma^\mathrm{2PI}}{\delta G}=0$
\begin{equation}
  \label{eq:Sigma}
  \Sigma(k) = 2\frac{i}{\hbar}\, 
  \frac{\delta \Gamma_2^\mathrm{2PI}[\phi,G]}{\delta G(k)}\ .
\end{equation}
So the 2PI self-energy is obtained by cutting a line of a 2PI 
vacuum graph
and considering a combinatorial factor. Again, the leading terms of the 
self-energy in a loop expansion are displayed in Fig.~\ref{fig:Sigma}.


\subsection{2PPI expansion}
The 2PPI effective action, 
proposed by Verschelde and Coppens~\cite{Verschelde:2PPI},
 is a variant of the 
``effective action of composite operators'' by 
Cornwall, Jackiw and Tomboulis~\cite{Cornwall:1974vz}.
We briefly repeat the formal derivation in Ref.~\cite{Verschelde:2PPI} 
without going too much into detail.

\begin{figure}[htbp]
  \centering
  \psfrag{G}{$\Gamma_q^\mathrm{2PPI}$}
  \psfrag{2PPI}{}
  \psfrag{2}{}
  \psfrag{q}{}
  \psfrag{=}{$=$}
  \psfrag{+}{$+$}
  \epsfig{file=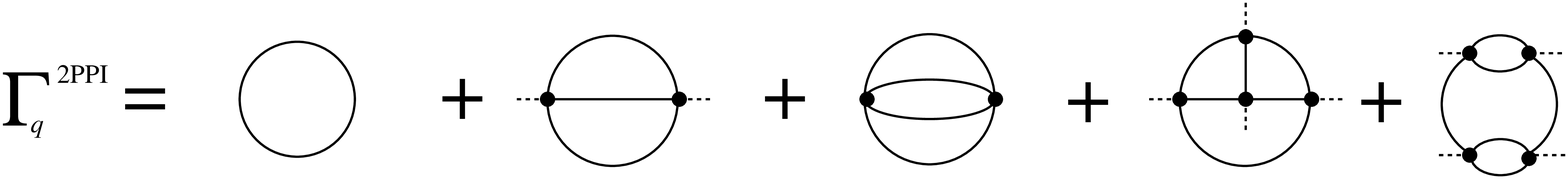,width=12cm}
  \caption{``Quantum'' part of the 2PPI effective action including some
    three-loop contributions. 
    Here, a line represents a \emph{bubble-resummed}, 
    i.e., local, propagator as defined in Eq.~(\ref{eq:2PPI_G}). 
    Note that the last graph does not appear in $\Gamma_2^\mathrm{2PI}$
    because it is 2PPI but 2PR.}
  \label{fig:2PPI}
\end{figure}

In the 2PI formalism one deals with a \emph{bilocal} composite
operator $\Phi(x)\, \Phi(y)$ which is coupled to an external
(bilocal) source $K(x,y)$, while
in the 2PPI approach one keeps this source \emph{local} by construction.
Using Euclidean space-time, as in Ref.~\cite{Verschelde:2PPI}, one
defines the following effective action of \emph{local} composite
operators
\begin{eqnarray}
  \Gamma^\mathrm{2PPI}[\phi,\Delta]&=& \mathcal{W}[J_1,J_2] - J_1\cdot\phi 
    - \frac{1}{2} J_2 \cdot (\phi^2+\Delta)
\end{eqnarray}
where
\begin{eqnarray}
  \frac{\delta\mathcal{W}}{\delta J_1} &=& \langle \Phi \rangle = \phi\\
  \frac{\delta\mathcal{W}}{\delta J_2} &=& \frac{1}{2}\langle \Phi^2 \rangle 
  = \frac{1}{2} (\phi^2+\Delta)
\end{eqnarray}
and
\begin{equation}
  \exp\{- \mathcal{W}[J_1,J_2]\} = \int\mathcal{D}\Phi\ \exp\{
    -( S[\Phi] + J_1\cdot\Phi + J_2\cdot\Phi^2) \}\ .
\end{equation}
The external fields (sources) $J_1$ and $J_2$ are both \emph{local} and 
fix the expectation
value of $\Phi$ and $\Phi^2$.
The effective equations of motion turn out to be
\begin{eqnarray}
  \frac{\delta\Gamma[\phi,\Delta]}{\delta\phi} &=& -J_1 - J_2 \phi\\
  \frac{\delta\Gamma[\phi,\Delta]}{\delta\Delta} &=& - \frac{1}{2}J_2\ .
\end{eqnarray}
We do not want to explain in detail the combinatorial trick used in 
Ref.~\cite{Verschelde:2PPI} for the derivative 
$\frac{\delta \Gamma^\mathrm{2PPI}[\phi,\Delta]}{\delta\phi}$
in order to sum all 2PR graphs. Eventually, the result is the 
complete effective action of the 2PPI formalism. It
consists of the classical action,
the ``quantum'' part and a constant which prevents double-counting:
\begin{equation}
  \label{eq:Gamma2PPI}
  \Gamma^\mathrm{2PPI}[\phi,\Delta]=
  S[\phi] + \Gamma_q^\mathrm{2PPI}[\phi,\mathcal{M}^2]
  -\frac{\lambda}{4} 3\,\lambda\Delta^2\ .
\end{equation}
The ``quantum'' part $\Gamma_q^\mathrm{2PPI}$ of the 2PPI action
contains the one-loop ``ln~det'' contribution and
all 2PPI graphs \emph{without} the ``double-bubble'' 
(see Fig.~\ref{fig:2PPI}). 
A \emph{two-particle-point-irreducible} (2PPI) graph is 1PI and 
stays connected whenever two internal lines meeting at the same point
(vertex) are cut \cite{Verschelde:2PPI}.
These graphs are to be computed using
a propagator with an effective mass $\mathcal{M}$
\begin{equation}
  \label{eq:2PPI_G}
  G(k) = \frac{1}{k^2+\mathcal{M}^2}\ .
\end{equation}
Since this propagator always remains local -- even if the \emph{exact}
2PPI effective action is computed -- 
it will never be equal to the \emph{physical} two-point Green function.

\begin{figure}[htbp]
  \centering
  \psfrag{D}{$\Delta$}
  \psfrag{=}{$=$}
  \psfrag{+}{$+$}
  \epsfig{file=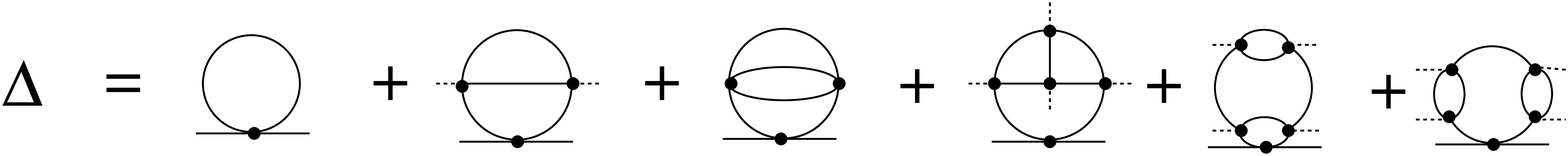,width=12cm}
  \caption{Contributions to the 2PPI self-energy with some three-loop
    contributions. 
    A line represents a local \emph{bubble-resummed}, 
    propagator as defined in Eq.~(\ref{eq:2PPI_G}). Note that
    the last two graphs do not appear in $\Gamma_2^\mathrm{2PI}$
    because they are 2PPI but 2PR.}
  \label{fig:Delta}
\end{figure}

The effective mass 
consists of the ``classical'' mass $(-\lambda v^2)$,
the seagulls $\lambda \phi^2$ and the local self-energy $\Delta$
\begin{equation}
  \label{eq:gap}
  \mathcal{M}^2 = \lambda (3\phi^2 -v^2) + 3\,\lambda\Delta\ .
\end{equation}
The self-energy is --- like in 2PI --- 
obtained as a derivative of the ``quantum''
part of the effective action
\begin{equation}
  \frac{\delta \Gamma_q^\mathrm{2PPI}[\phi,\mathcal{M}^2]}
    {\delta \mathcal{M}^2}
   = \frac{1}{2} \Delta\ .
\end{equation}
 This is equivalent to cutting a line (by
deriving with respect to the propagator $G(k)$) and then connecting
the two endpoints to a common third point by considering the inner derivative
\begin{equation}
\frac{\delta}{\delta \mathcal{M}^2} = \int [dk]\ \frac{\delta G(k)}
   {\delta \mathcal{M}^2} \frac{\delta}{\delta G(k)}
   = \int [dk]\ \frac{-1}{(k^2+\mathcal{M}^2)^2} 
     \frac{\delta}{\delta G(k)}
   \ .  
\end{equation}
The first (one-loop) ``quantum correction'' to the 2PPI effective action
only consists of the ``ln~det'' term in $\Gamma_q^\mathrm{2PPI}$. 
This approximation is equivalent to what is called ``Hartree'' in 2PI, 
namely the resummation of tadpoles or daisy and super-daisy graphs. The first
two-loop contribution to the 2PPI effective action is the sunset diagram.
The mass corrections resulting from all one-, two- and three-loop 
vacuum diagrams of Fig.~\ref{fig:2PPI}
are shown in Fig.~\ref{fig:Delta}.

Now, we use Eq.~\eqn{eq:gap} to express $\Delta$ in terms of $\mathcal{M}^2$
and insert this expression into the effective action~\eqn{eq:Gamma2PPI}.
We finally obtain the 2PPI effective action in terms of the mean
field $\phi$ and the effective mass $\mathcal{M}$. Restricting
to a homogeneous mean field and using the explicit
form of the classical Lagrangian~\eqn{eq:Lagrange} the 
effective potential reads
\begin{eqnarray}
  \nonumber
  V_\mathrm{eff}(\phi,\mathcal{M}^2) &=& 
     \frac{1}{2} \mathcal{M}^2
      \phi^2 - \frac{\lambda}{2} \phi^4 \\
   &&  + V^\mathrm{2PPI}_q(\phi,\mathcal{M}^2)
     - \frac{1}{12\,\lambda} \left[ \mathcal{M}^2 + \lambda v^2 
       \right]^2\ .
\end{eqnarray}
This is the effective potential from which the equations of motion can be
(re)-obtained by varying $\phi$ and $\calm^2$. 
This procedure can be generalized to obtain an effective action for 
nonequilibrium dynamics and a conserved energy functional
 \cite{Baacke:2002ee}.




\section{The sunset diagram at $T=0$}
\setcounter{equation}{0}
The sunset diagram at $T=0$ has been evaluated previously by several
authors \cite{vanderBij:1983bw,Davydychev:1992mt}. Here we are interested in this diagram
with external momentum zero, and with at least two internal lines
of equal mass. We give here some technical details.
\subsection{The $\sigma\sigma\sigma$ diagram}
The sunset diagram for equal masses has been given in \cite{vanderBij:1983bw}.
The authors use $d=4+\epsilon$ and omit the factor $1/(2\pi)^d$.
Alternatively one may use the expression given in \cite{Davydychev:1992mt}
for different masses and set all masses equal. These authors use
$d=4-2\epsilon$ and likewise omit the factor $1/(2\pi)^d$.
Accomodating both expressions to our standard
\be
I_{\sigma\sigma\sigma}=\int\frac{d^{4-\epsilon}p}{(2\pi)^{4-\epsilon}}
\int\frac{d^{4-\epsilon}q}{(2\pi)^{4-\epsilon}}
\frac{1}{(p^2+M^2)(q^2+M^2)((p-q)^2+M^2)}
\ee
we find
\bea\nonumber
I_{\sigma\sigma\sigma}&=&
-\frac{3 M^2}{(4\pi)^4}\left\{
\frac{2}{\epsilon^2}+\frac{2}{\epsilon}\left(\ln\frac{4\pi\mu^2}{M^2}-\gamma_E
+\frac{3}{2}\right)\right.
\\&&\left.+\left(\ln\frac{4\pi\mu^2}{M^2}-\gamma_E
+\frac{3}{2}\right)^2+\frac{\pi^2}{12}+\frac{5}{4}
  -\frac{2}{\sqrt{3}}{\rm Cl}\left(\frac{\pi}{3}\right)\right\}
\kma\eea
where ${\rm Cl}(\phi)$ is the Clausen function
\be
 {\rm Cl}(\varphi)=\sum_{k=1}^{\infty}\frac{\sin k \varphi}{k^2}
\pkt\ee

\subsection{The $\sigma\pi\pi$ diagram}
Defining the sunset integral for one $\sigma$ line with mass $M=m_\sigma$
and two pion lines with mass $m=m_\pi$ as
\be
I_{\sigma\pi\pi}=\int\frac{d^{4-\epsilon}p}{(2\pi)^{4-\epsilon}}
\int\frac{d^{4-\epsilon}q}{(2\pi)^{4-\epsilon}}
\frac{1}{(p^2+M^2)(q^2+m^2)((p-q)^2+m^2)}
\kma \ee
we again use the expressions given by Davydychev and Tausk
\cite{Davydychev:1992mt} which we have cross-checked with the
the reduction formula of Ref. \cite{vanderBij:1983bw}. When adapted to our
conventions we obtain
\bea
I_{\sigma\pi\pi}&=&\nonumber
-\frac{2 m^2 (1+2z)}{(4\pi)^4}\left\{
\frac{2}{\epsilon^2}+\frac{2}{\epsilon}\left(\ln\frac{4\pi\mu^2}{M^2}-\gamma_E
+\frac{3}{2}-\frac{2z}{1+2z}\ln 4z\right)\right.
\\&&+\left(\ln\frac{4\pi\mu^2}{M^2}-\gamma_E \nonumber
+\frac{3}{2}-\frac{2z}{1+2z}\ln 4z\right)^2+\frac{\pi^2}{12}+\frac{5}{4}
\\
&&\left.+\frac{z(1-4z)}{1+2z}\ln^2 4z
 -\frac{4\sqrt{z(1-z)}}{1+2z}{\rm Cl}(\varphi)\right\}
\eea
where $z=M^2/4 m^2$ and $\varphi=\arccos (1-2z)$.

\section{The sunset diagram with one thermal line}
\setcounter{equation}{0}
If one of the lines of the sunset diagram is replaced by a thermal line
it takes the form
\be
I^\beta_{ij|\underline{k}}=\int\frac{d^{3-\epsilon}p}{(2\pi)^{3-\epsilon}E_k(\bfp)}
n_k^\beta(\bfp)\int\frac{d^{4-\epsilon}q}{(2\pi)^{4-\epsilon}}
\frac{1}{(q^2+m_i^2)((p-q)^2+m_j^2)}
\pkt\ee
Here $E_k(\bfp)=\sqrt{\bfp^2+m_k^2}$ and $n^\beta_k(\bfp)$ is
the Bose-Einstein distribution function
\be
n^\beta_k(\bfp)=\frac{1}{\exp(E_k/T)-1}
\pkt\ee
The second integral is the fish diagram with the external
Euclidean momentum $p$ which is on shell, i.e., $p^2=-m_k^2$
\be
F(m_i,m_j;m_k)=\int\frac{d^{4-\epsilon}q}{(2\pi)^{4-\epsilon}}
\frac{1}{(q^2+m_i^2)((p-q)^2+m_j^2)}\biggr|_{p^2=-m_k^2}
\pkt\ee
Explicitly one finds
\bea
F(m_i,m_j;m_k)&=&\frac{1}{(4\pi)^2}\left\{\frac{2}{\epsilon}-\gamma_E
+ \ln 4 \pi - \ln \frac{m_k^2}{\mu^2}
\right . \nonumber
\\\label{eq:fishalpha}
&& \left.-\int_0^1 d\alpha \ln \left[\alpha \frac{m_i^2}{m_k^2} +(1-\alpha)
\frac{m_j^2}{m_k^2}-\alpha(1-\alpha)\right]\right\}
\pkt\eea
The $\alpha$ integration can be done analytically, leading to
various expressions in terms of logarithms or
inverse trigonometric functions, depending on the relations
among the three masses. We note in particular that for
$m_k > m_i+m_j$ the integral (\ref{eq:fishalpha}) develops an imaginary part.
In the present context this happens for $m_k=m_\sigma$ and
$m_i=m_j=m_\pi< m_\sigma/2$. The thermal integral
$I^\beta_{ij|\underline k}$ thereby
gets an imaginary part as well, reflecting the fact that
in the heat bath the sigma particles can decay into or be produced
by pions.
The sunset integral with one thermal line factorizes:
\be
I^\beta_{ij|\underline{k}}=
F(m_i,m_j;m_k)I^{\beta}_{3-\epsilon}(m_k)
\ee
with
\be
I^{\beta}_{3-\epsilon}(m_k)=
\int\frac{d^{3-\epsilon}p}{(2\pi)^{3-\epsilon}E_k(\bfp)}
n_k^\beta(\bfp)
\pkt\ee
The divergent part obviously takes the form
\be
I^{\beta,div}_{ij|\underline k}=\frac{1}{(4\pi)^2}\frac{2}{\epsilon}
I^{\beta}_{3-\epsilon}(m_k)
\pkt
\ee
The general expression agrees with Ref. \cite{Parwani:1991gq}, Eq. (3.7),
we use a different regularization, however.

\section{The sunset diagram with two thermal lines}
\setcounter{equation}{0}
We define the sunset diagram with two thermal lines as
\be
I^\beta_{i|\underline{jk}}=
\int\frac{d^{3-\epsilon}p}{(2\pi)^{3-\epsilon}2E_k(\bfp)}
n_k^\beta(\bfp)\int\frac{d^{3-\epsilon}q}{(2\pi)^{3-\epsilon}2E_j(\bfq)}
n_j^\beta(\bfq)
\sum_{r,s=\pm}\frac{1}{(p_r+q_s)^2+m_i^2}
\pkt\ee
Here $p_r$ and $q_s$ are Euclidean momenta with
$p_\pm=\pm(iE_k(\bfp),\bfp)$ and $q_\pm=\pm(iE_j(\bfq),\bfq)$.
The integration over the angle between $\bfp$ and $\bfq$ can be
done analytically, with the result~\cite{Parwani:1991gq}
\be
I^\beta_{i|\underline{jk}}=
\frac{1}{32 \pi^4}\int_0^\infty\frac{p\;dp}{E_k(\bfp)}n^\beta_k(\bfp)
\int_0^\infty\frac{q\;dq}{E_j(\bfq)}n^\beta_j(\bfq)
\ln\biggl|\frac{Y_+}{Y_-}\biggr|
\ee
where
\be
Y_\pm=\left\{[E_k(p)+E_j(q)]^2-E^2_i(p\pm q)\right\}
\left\{[E_k(p)-E_j(q)]^2-E^2_i(p\pm q)\right\}
\pkt
\ee
The integrand has logarithmic singularities within the
region of integration which have to be treated
with care in the numerical integration over $p=|\bfp|$ and
$q=|\bfq|$.

\end{appendix}




\end{document}